\documentclass[twocolumn,pre,superscriptaddress,longbibliography]{revtex4}
\usepackage{graphics}
\usepackage{graphicx}
\usepackage{amsfonts}
\usepackage{textcomp}
\usepackage{amssymb}
\usepackage{mathrsfs}
\usepackage{amsmath}
\usepackage{color}
\usepackage{ulem}
\begin{document}
\title{Vibrational Properties of One-Dimensional Disordered Hyperuniform Atomic Chains}

\author{Houlong Zhuang}
\email[correspondence sent to: ]{hzhuang7@asu.edu}
\affiliation{School for Engineering of Matter, Transport and
Energy, Arizona State University, Tempe, AZ 85287}
\author{Duyu Chen}
\affiliation{Materials Research Laboratory, University of California, Santa Barbara, California 93106, United States}
\author{Lei Liu}
\affiliation{School for Engineering of Matter, Transport and
Energy, Arizona State University, Tempe, AZ 85287}
\author{Ge Zhang}
\affiliation{Department of Physics, City University of Hong Kong, Hong Kong, China}
\author{Yang Jiao}
\email[correspondence sent to: ]{yang.jiao.2@asu.edu}
\affiliation{Materials Science and Engineering, Arizona State
University, Tempe, AZ 85287} \affiliation{Department of Physics,
Arizona State University, Tempe, AZ 85287}


\begin{abstract}
Disorder hyperuniformity (DHU) is a recently discovered exotic state of many-body systems that possess a hidden order in between that of a perfect crystal and a completely disordered system. Recently, this novel DHU state has been observed in a number of quantum materials including amorphous 2D graphene and silica, which are endowed with unexpected electronic transport properties. Here, we numerically investigate 1D atomic chain models, including perfect crystalline, disordered hyperuniform as well as randomly perturbed atom packing configurations to obtain a quantitative understanding of how the unique DHU disorder affects the vibrational properties of these low-dimensional materials. We find that the DHU chains possess lower cohesive energies compared to the randomly perturbed chains, implying their potential reliability in experiments. Our inverse partition ratio (IPR) calculations indicate that the DHU chains can support fully delocalized states just like perfect crystalline chains over a wide range of frequencies, i.e., $\omega \in (0, 100)$ cm$^{-1}$, suggesting superior phonon transport behaviors within these frequencies, which was traditionally considered impossible in disordered systems. Interestingly, we observe the emergence of a group of highly localized states associated with $\omega \sim 200$ cm$^{-1}$, which is characterized by a significant peak in the IPR and a peak in phonon density of states at the corresponding frequency, and is potentially useful for decoupling electron and phonon degrees of freedom. These unique properties of DHU chains have implications in the design and engineering of novel DHU quantum materials for thermal and phononic applications.
\end{abstract}


\maketitle



\section{Introduction}

Disorder hyperuniformity (DHU) is a recently discovered exotic state of many-body systems \cite{To03, To18a}. DHU systems are statistically isotropic and lack traditional long-range orders as in an amorphous material, yet they suppress large-scale normalized density fluctuations like crystals \cite{To03}. DHU is characterized by a local number variance $\sigma_N^2(R)$ associated with a spherical window of radius $R$ that grows more slowly than
the window volume (e.g., with scaling $R^d$ in $d$-dimensional
Euclidean space) in the large-$R$ limit \cite{To03, To18a}, i.e.,
\begin{equation}
\lim_{r\rightarrow \infty} \sigma_N^2(R)/R^d = 0.
\end{equation}
This is equivalently manifested as the
vanishing static structure factor in the infinite-wavelength (or zero-wavenumber) limit, i.e., $\lim_{k\rightarrow 0}S(k) = 0$, where $k$ is the wavenumber. The small-$k$ scaling behavior of $S(k) \sim k^\alpha$ determines the large-$R$
asymptotic behavior of $\sigma_N^2(R)$, based on which all DHU
systems can be categorized into three classes:
$\sigma_N^2(R) \sim R^{d-1}$ for $\alpha>1$ (class I); $\sigma_N^2(R)
\sim R^{d-1}\ln(R)$ for $\alpha=1$ (class II); and $\sigma_N^2(R)
\sim R^{d-\alpha}$ for $0<\alpha<1$ (class III) \cite{To18a}.


A wide spectrum of equilibrium and non-equilibrium many-body systems, in both classical and quantum mechanical settings, have been identified to possess the property of
hyperuniformity. Examples include density fluctuations in early universe \cite{Ga02}, jammed hard-particle packings \cite{Do05, Za11a, zachary2011hyperuniformity, Ji11, yuan2021universality}, exotic classical \cite{To15, Uc04, Ba08, Ba09, jiao2022hyperuniformity} and quantum ground states \cite{To08, Fe56}, jammed colloidal systems \cite{Ku11, Hu12, Dr15}, driven non-equilibrium systems \cite{He15, Ja15, We15, salvalaglio2020hyperuniform, nizam2021dynamic, zheng2023universal}, active fluids \cite{Le19, lei2019hydrodynamics, huang2021circular, zhang2022hyperuniform, oppenheimer2022hyperuniformity}, dynamic random organizing systems
\cite{hexner2017noise, hexner2017enhanced, weijs2017mixing,
wilken2022random}, biological systems across scales \cite{Ji14, Ma15, ge2023hidden}, vortex distribution in superconductors \cite{Ru19, Sa19}, electron density distributions \cite{Ge19quantum, sakai2022quantum}, as well as a wide class of disordered materials \cite{chen2023disordered}, including amorphous 2D materials \cite{Zh20, Ch21, PhysRevB.103.224102, Zh21}, medium/high entropy alloys \cite{chen2021multihyperuniform}, defected carbon nano-tubes \cite{nanotube}, certain metallic glasses \cite{zhang2023approach}, and polymeric materials \cite{ Ch18b}.


Many DHU materials are found to possess
superior physical properties including large isotropic photonic band gaps \cite{Fl09, Ma13, klatt2022wave}, optimized transport properties \cite{Zh16, Ch18a, torquato2021diffusion}, mechanical properties \cite{Xu17}, wave-propagation characteristics \cite{Ch18a, Kl18, Le16, kim2023effective}, scattering properties \cite{yu2023evolving, shi2023computational}, as well as optimal multi-functionalities \cite{To18b}. Very recently, quasi-1D and 1D DHU systems have been investigated, including defected nanotubes \cite{nanotube} as well as certain 1D quasi-periodic models including the tight-binding Fibonacci models and the Aubry-Andre-Harper (AAH) model \cite{sakai2022quantum}. In the latter case, it was found that the electron densities induced by these quasi-periodic model Hamiltonians were hyperuniform. These exciting discoveries not only suggest the existence of a novel DHU state of electrons in low dimensional materials, but also shed lights on novel device applications by exploring the unique emergent properties of the DHU electron
states.


In this work, we focus on disordered hyperuniform 1D atomic chains. Despite the fact that 1D atomic chains is generally not energetically stable and difficult to realize experimentally, the electron and phonon behavior of various 1D systems has been the subject of interest for theoretical physicists for several decades. Many fascinating physical phenomena occurs as a result of the (structural or compositional) disorder. One notable example was the electron localization predicted by Anderson in the 1950s \cite{anderson1958absence}. Many previous theoretical studies \cite{czycholl1980numerical, kramer1993localization} were based on a tight-binding Hamiltonian model. Although more advanced theoretical tools such as Hohenberg-Kohn-Sham density functional theory (DFT) \cite{hohenberg1964inhomogeneous, kohn1965self} have been developed over the last several decades, 1D tight-binding model has also been vastly used in the literature to derive many interesting physical properties. For example, Hoffman et al. used crystalline hydrogen and carbon atomic chains of various spacings as examples to illustrate the tight-binding method \cite{hoffmann1991chemical}. One electron in a 1D periodic potential can be described as the Bloch waves extended in the whole crystal. It remains unknown what is the difference between the electronic structure of a 1D DHU system and its crystalline or random counterparts.

In this work, we investigate how the correlated disordered induced via hyperuniformity influences the behaviors of electrons and phonons in model 1D atomic chains, compared to both the crystalline and uncorrelated disordered counterparts. In particular, we find the DHU chains possess lower cohesive energies compared to the randomly perturbed chains, implying their potential reliability in experiments. We also compute the inverse partition ratio (IPR) for the vibrational states and find that the DHU chains can support fully delocalized states over a wide range of frequencies, i.e., $\omega \in (0, 100)$ cm$^{-1}$, indicated by the minimal values of IPR within this range. This result suggests superior phonon transport behaviors within these frequencies, which was traditionally considered not possible in disordered systems. Moreover, we observe the emergence of a group of highly localized states associated with $\omega \sim 200$ cm$^{-1}$, which is characterized by a significant peak in the IPR graph and a peak in density of states (DOS) at the corresponding frequency. These unique properties of DHU chains have implications in the design and engineering of novel DHU quantum materials for thermal and phononic applications.



\section{Model 1D Atomic Chains}

\begin{figure}[ht!]
\begin{center}
$\begin{array}{c}\\
\includegraphics[width=0.485\textwidth]{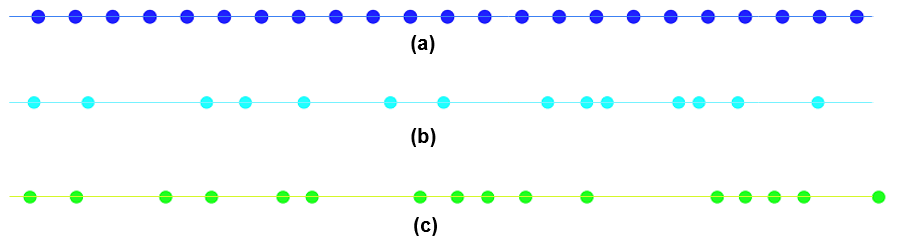}
\end{array}$
\end{center}
\caption{Representative configurations of different types of 1D atomic chains: (a) crystalline, (b) randomly perturbed, (c) stealthy hyperuniform chains. The configurations used in the simulations contain $N = 6766$ particles and only a small portion of the full configuration is shown here for each system.} \label{fig_1}
\end{figure}

\begin{figure}[ht!]
\begin{center}
$\begin{array}{c}\\
\includegraphics[width=0.485\textwidth]{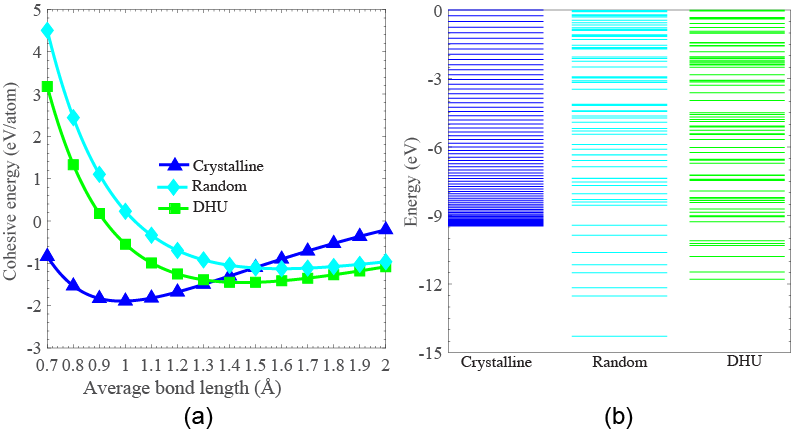}
\end{array}$
\end{center}
\caption{(a) Cohesive energies of the three atomic hydrogen chains as a function of average bond length. (b) Eigenvalue distributions of the atomic hydrogen chains.} \label{fig_2}
\end{figure}

\begin{figure*}[ht!]
\begin{center}
$\begin{array}{c}\\
\includegraphics[width=0.685\textwidth]{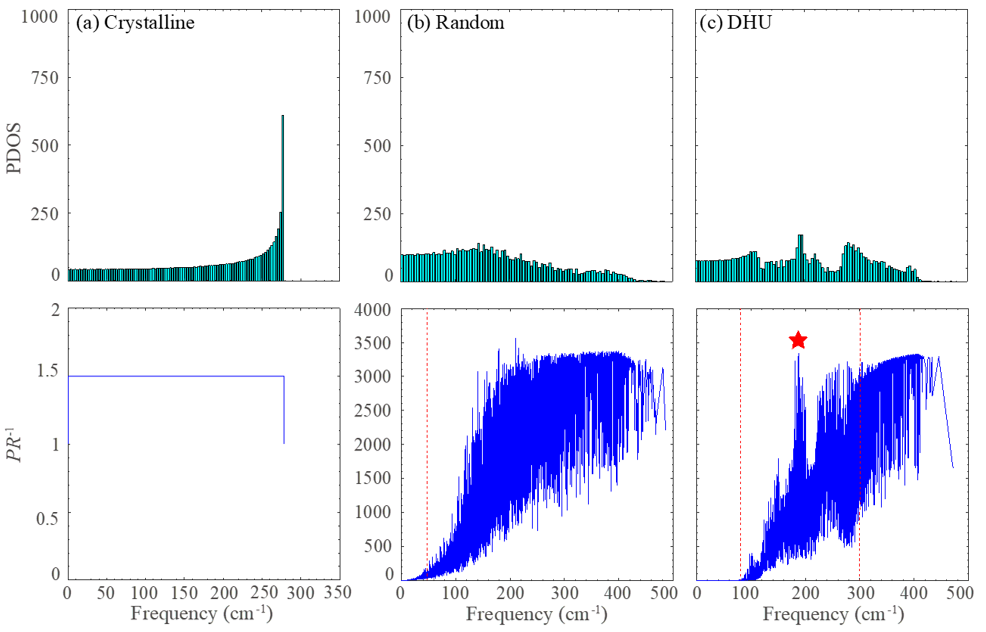}
\end{array}$
\end{center}
\caption{Phonon density of states (PDOS, top panels) and the inverse partition ratio ($PR^{-1}$, bottom panels) of 1D (a) crystalline, (b) randomly perturbed, and (c) DHU carbon atomic chains. The red dashed line in (b) indicates the mobility edge. The red dashed lines in (c) encloses a peak (denoted by a red star) in IPR that is a unique feature of DHU state.} \label{fig_3}
\end{figure*}

We start with obtaining numerical solutions for various model systems of 1D atomic chains. Specifically, an atomic chain is a 1D packing of atoms, typically on a substrate, where the distance between each pair of successive atoms can be prescribed in principle. Although first introduced as computational model, these atomic chains are not completely unrealistic. On the contrary, it has been found in experiment that 1D crystalline atomic chain of significant length has been successfully synthesized \cite{curl1988probing}.

In this work, we consider three types of atomic chains:
\begin{itemize}
    \item (i) a crystalline system based on simple integer lattice in 1D where the inter-atomic spacing $\ell$ is equal;
    \item (ii) uncorrelated disordered systems obtained by perturbing the positions of the atoms from an integer lattice with uniform displacement $\delta \in [-0.1 \ell, 0.1 \ell]$;
    \item (iii) DHU systems where the atomic centers correspond to a stealthy hyperuniform distribution (i.e., with $S(k) = 0$ for $k < K^*$) generated using the collective coordinate method \cite{Ba08}. The choice of $K^*$ leads to different numbers of constraints imposed, and the parameter $\chi$ characterizes the fraction of constraints over the total degrees of freedom \cite{To15}. Here, we use $\chi = 0.3$ for which the associated systems still possess a relatively high degree of disorder (see Fig. 1(c)).
\end{itemize}


Without loss of generality, we use atomic chains of hydrogen and carbon atoms with a total number $N$ = 145 and 6766 as examples here for DFT and model Hamiltonian calculations, respectively. We select the hydrogen chain for the eigenvalue plot because it has fewer occupied energy levels, ensuring a clearer presentation in the resulting plot. For vibrational frequency calculations, we apply the calculations to carbon chains since the number of atoms that can be simulated is not limited by the capabilities of DFT calculations. The representative configurations for the three types of chains are shown in Fig. 1. For DFT calculations, we use the Vienna Ab initio Simulation Package (VASP) \cite{kresse1996efficiency,PhysRevB.54.11169} with the computational parameters: a cutoff kinetic energy of 600 eV, an energy convergence criterion of $10^{-6}$ eV, a single Gamma point, and the standard Perdew-Burke-Ernzerhof (PBE)\cite{perdew1996generalized} potential from the projector augmented wave (PAW) method \cite{PhysRevB.59.1758}.

Similar to the tight-binding model Hamiltonian, a dynamical matrix for a 1D atomic chain can be constructed considering only the nearest-neighbor interactions, which allows to obtain vibrational properties of these systems. Each matrix element is related to distance-dependent spring constant $k_{ij}$ \cite{zhou2016phonon}, i.e.,
\begin{equation}
k_{ij} = k_0 ({r_{ij}}/a_0)^{-2},
\end{equation}
where $k_0$ is the spring constant, $r_{ij}$ is the inter-atomic spacing between atoms $i$ and $j$, and $a_0$ is characteristic length scale (e.g., the lattice constant for a 1D crystalline chain; 1.27 \AA ~ for the carbon chains). The periodic boundary conditions are applied along the chains. $k_0$ (55.58 eV/\AA$^2$) is determined by performing DFT energy calculations on a diatomic molecule with different $r_{ij}$.

\section{Energetics and Vibrational Properties of 1D Atomic Chains}

We first compute the minimum cohesive energy of different hydrogen chains, for which energies of isolated atoms are used as the reference and the results are shown in Fig. 2(a).  It can be seen that the minimum cohesive energy of 1D DHU hydrogen chain lies between the minimum of 1D crystalline and amorphous atomic chains, which implies these DHU chains can be potentially realized in experiments. In addition, these results reveal that the energy (eigenvalue) distribution of the three systems are quantitatively different. In the 1D crystalline system, the eigenvalues (see Fig. 2(b)) are roughly separated into two groups at high and low energies, where the distributions of the energy levels are homogeneous. In the randomly perturbed system, the energy levels are spread out to the most extent. In the 1D DHU system, the distribution is between the crystalline and random cases.

To analyze the vibrational properties of the 1D atomic chains, we diagonalize the associated dynamic matrix to compute the eigenvalue and the corresponding eigenvectors to obtain the numerical solutions. We focus here on the phonon vibrational frequencies and their corresponding normal modes (eigenvectors). Since phonons are one main carrier for thermal transport, we also apply metrics to analyze the phonon eigenvector.
Specifically, we examine the phonon localization by computing the inverse participation ratio ($PR^{-1}_n$) for each vibrational mode n using the following equation \cite{lee2017molecular, guo2015approaching}:
\begin{equation}
PR^{-1}_n = N \sum_i e^4_{i,n}/(\sum_i e^2_{i,n})^2,
\end{equation}
where N is the total number of atoms and $e_{i,n}$ is the phonon eigenvector of mode n. High and low $PR^{-1}$ values correspond to small and large eigenvectors, respectively.

Figure 3 shows the phonon density of states (PDOS) (upper panels) and inverse partition ratio (lower panels) of the three different 1D atomic chains. For the crystalline chain, we observe that the well-known analytical solution is well reproduced \cite{grosso2003giuseppe}. In particular, a singularity at the highest phonon frequency is very apparent. The inverse partition ratio $PR^{-1}$ is constant for all frequencies and close to zero, as the phonon modes are similar to the Bloch’s wave and thus, are fully delocalized spatially. For the randomly perturbed chain, we observe the mobility edge (indicated as the red dashed line in Fig. 3b) between two groups of $PR^{-1}$. According to Allen et al. \cite{allen1999diffusons}, the phonon modes below the mobility edge is classified as ``locons'', which can be considered as the phonon version of Anderson localization. The mobility edge that separates ``locons'' from the so-called ``propagons'' and ``diffusons'' that are delocalized states positively contributing to the thermal conductivity of the material.

The inverse partition ratio $PR^{-1}$ for the DHU chain exhibits peculiar features, clearly demonstrating the effects of the unique disorder on the material properties. First, we observe some similarity in the $PR^{-1}$ to the crystalline atomic chain in the low-frequency region ($\omega <$ 100 cm$^{-1}$); and to the randomly perturbed chain in the high-frequency region ($\omega >$ 300 cm$^{-1}$). Specifically, the $PR^{-1}$ of the DHU chain is constant and close to zero for $\omega <$ 100 cm$^{-1}$, which indicates DHU chains can support fully delocalized states (i.e., Bloch's waves) and thus, thermal transport for this wide frequency range. This was considered not possible in 1D disordered systems in conventional wisdom. On the other hand, the $PR^{-1}$ of the DHU chain almost reaches the maximal value as in the randomly perturbed chain for $\omega >$ 300 cm$^{-1}$, indicating localized states similar to ``locons'' for these frequencies. Notably, we observe a significant peak in the $PR^{-1}$ of the DHU chain around the intermediate frequencies ($\approx$ 200 cm$^{-1}$), which also corresponds to a strong peak in PDOS. These peaks indicate the emergence of a large number of highly localized states in DHU chains at these intermediate frequencies.



\section{Conclusions and Discussion}

To summarize, we investigated the vibrational properties of model 1D DHU atomic chains in this work, by comparing the phonon density of states and inverse partition ratio of the system to the crystalline and randomly perturbed chains. We observed two unique properties resulted from the hyperuniform nature of the system. The first one is the support of fully delocalized states and thus thermal transport for a wide range of frequencies in the disordered hyperuniform chains, which was considered to be impossible in conventional wisdom. This property would allow the design and engineering of novel DHU materials with superior thermal transport properties that are highly robust to defects. The second one is the emergence of a group of highly localized states in the intermediate frequencies. This would allow one to design novel quantum materials that decouple electron and phonon degrees of freedom for quantum decoherence protection for quantum computing applications.

There are also many interesting open questions that need to be addressed. For example, it is not completely understood how the DHU correlations led to the observed peak in the $PR^{-1}$ and the emergence of the associated highly localized states at these frequencies. Insights might be obtained by analyzing the displacement patterns that are associated with the peak. In addition, it is unclear if the frequency associated with peak can be engineered by varying the degree of order and correlations in the DHU chains. We will explore these open questions in our future work. Finally, we note that although we only focused on 1D atomic chains here, the insights obtained would also be applicable in the vibrational properties of 2D DHU quantum materials.





\begin{acknowledgments}
H.Z. thank the start-up funds from ASU. This research used computational resources of the Agave Research Computer Cluster of ASU and the Texas Advanced Computing Center under Contract No. TG-DMR170070. Y. J. is supported by the Army Research Office under Cooperative Agreement No. W911NF-22-2-0103.
\end{acknowledgments}

\bibliography{network}

\begin{thebibliography}{78}
\expandafter\ifx\csname natexlab\endcsname\relax\def\natexlab#1{#1}\fi
\expandafter\ifx\csname bibnamefont\endcsname\relax
  \def\bibnamefont#1{#1}\fi
\expandafter\ifx\csname bibfnamefont\endcsname\relax
  \def\bibfnamefont#1{#1}\fi
\expandafter\ifx\csname citenamefont\endcsname\relax
  \def\citenamefont#1{#1}\fi
\expandafter\ifx\csname url\endcsname\relax
  \def\url#1{\texttt{#1}}\fi
\expandafter\ifx\csname urlprefix\endcsname\relax\def\urlprefix{URL }\fi
\providecommand{\bibinfo}[2]{#2}
\providecommand{\eprint}[2][]{\url{#2}}

\bibitem[{\citenamefont{Torquato and Stillinger}(2003)}]{To03}
\bibinfo{author}{\bibfnamefont{S.}~\bibnamefont{Torquato}} \bibnamefont{and}
  \bibinfo{author}{\bibfnamefont{F.~H.} \bibnamefont{Stillinger}},
  \bibinfo{journal}{Phys. Rev. E} \textbf{\bibinfo{volume}{68}},
  \bibinfo{pages}{041113} (\bibinfo{year}{2003}).

\bibitem[{\citenamefont{Torquato}(2018)}]{To18a}
\bibinfo{author}{\bibfnamefont{S.}~\bibnamefont{Torquato}},
  \bibinfo{journal}{Phys. Rep.} \textbf{\bibinfo{volume}{745}},
  \bibinfo{pages}{1} (\bibinfo{year}{2018}).

\bibitem[{\citenamefont{Gabrielli et~al.}(2002)\citenamefont{Gabrielli, Joyce,
  and Labini}}]{Ga02}
\bibinfo{author}{\bibfnamefont{A.}~\bibnamefont{Gabrielli}},
  \bibinfo{author}{\bibfnamefont{M.}~\bibnamefont{Joyce}}, \bibnamefont{and}
  \bibinfo{author}{\bibfnamefont{F.~S.} \bibnamefont{Labini}},
  \bibinfo{journal}{Phys. Rev. D} \textbf{\bibinfo{volume}{65}},
  \bibinfo{pages}{083523} (\bibinfo{year}{2002}).

\bibitem[{\citenamefont{Donev et~al.}(2005)\citenamefont{Donev, Stillinger, and
  Torquato}}]{Do05}
\bibinfo{author}{\bibfnamefont{A.}~\bibnamefont{Donev}},
  \bibinfo{author}{\bibfnamefont{F.~H.} \bibnamefont{Stillinger}},
  \bibnamefont{and} \bibinfo{author}{\bibfnamefont{S.}~\bibnamefont{Torquato}},
  \bibinfo{journal}{Phys. Rev. Lett.} \textbf{\bibinfo{volume}{95}},
  \bibinfo{pages}{090604} (\bibinfo{year}{2005}).

\bibitem[{\citenamefont{Zachary
  et~al.}(2011{\natexlab{a}})\citenamefont{Zachary, Jiao, and
  Torquato}}]{Za11a}
\bibinfo{author}{\bibfnamefont{C.~E.} \bibnamefont{Zachary}},
  \bibinfo{author}{\bibfnamefont{Y.}~\bibnamefont{Jiao}}, \bibnamefont{and}
  \bibinfo{author}{\bibfnamefont{S.}~\bibnamefont{Torquato}},
  \bibinfo{journal}{Phys. Rev. Lett.} \textbf{\bibinfo{volume}{106}},
  \bibinfo{pages}{178001} (\bibinfo{year}{2011}{\natexlab{a}}).

\bibitem[{\citenamefont{Zachary
  et~al.}(2011{\natexlab{b}})\citenamefont{Zachary, Jiao, and
  Torquato}}]{zachary2011hyperuniformity}
\bibinfo{author}{\bibfnamefont{C.~E.} \bibnamefont{Zachary}},
  \bibinfo{author}{\bibfnamefont{Y.}~\bibnamefont{Jiao}}, \bibnamefont{and}
  \bibinfo{author}{\bibfnamefont{S.}~\bibnamefont{Torquato}},
  \bibinfo{journal}{Physical Review E} \textbf{\bibinfo{volume}{83}},
  \bibinfo{pages}{051309} (\bibinfo{year}{2011}{\natexlab{b}}).

\bibitem[{\citenamefont{Jiao and Torquato}(2011)}]{Ji11}
\bibinfo{author}{\bibfnamefont{Y.}~\bibnamefont{Jiao}} \bibnamefont{and}
  \bibinfo{author}{\bibfnamefont{S.}~\bibnamefont{Torquato}},
  \bibinfo{journal}{Phys. Rev. E} \textbf{\bibinfo{volume}{84}},
  \bibinfo{pages}{041309} (\bibinfo{year}{2011}).

\bibitem[{\citenamefont{Yuan et~al.}(2021)\citenamefont{Yuan, Jiao, Wang, and
  Li}}]{yuan2021universality}
\bibinfo{author}{\bibfnamefont{Y.}~\bibnamefont{Yuan}},
  \bibinfo{author}{\bibfnamefont{Y.}~\bibnamefont{Jiao}},
  \bibinfo{author}{\bibfnamefont{Y.}~\bibnamefont{Wang}}, \bibnamefont{and}
  \bibinfo{author}{\bibfnamefont{S.}~\bibnamefont{Li}},
  \bibinfo{journal}{Physical Review Research} \textbf{\bibinfo{volume}{3}},
  \bibinfo{pages}{033084} (\bibinfo{year}{2021}).

\bibitem[{\citenamefont{Torquato et~al.}(2015)\citenamefont{Torquato, Zhang,
  and Stillinger}}]{To15}
\bibinfo{author}{\bibfnamefont{S.}~\bibnamefont{Torquato}},
  \bibinfo{author}{\bibfnamefont{G.}~\bibnamefont{Zhang}}, \bibnamefont{and}
  \bibinfo{author}{\bibfnamefont{F.~H.} \bibnamefont{Stillinger}},
  \bibinfo{journal}{Phys. Rev. X} \textbf{\bibinfo{volume}{5}},
  \bibinfo{pages}{021020} (\bibinfo{year}{2015}).

\bibitem[{\citenamefont{Uche et~al.}(2004)\citenamefont{Uche, Stillinger, and
  Torquato}}]{Uc04}
\bibinfo{author}{\bibfnamefont{O.~U.} \bibnamefont{Uche}},
  \bibinfo{author}{\bibfnamefont{F.~H.} \bibnamefont{Stillinger}},
  \bibnamefont{and} \bibinfo{author}{\bibfnamefont{S.}~\bibnamefont{Torquato}},
  \bibinfo{journal}{Phys. Rev. E} \textbf{\bibinfo{volume}{70}},
  \bibinfo{pages}{046122} (\bibinfo{year}{2004}).

\bibitem[{\citenamefont{Batten et~al.}(2008)\citenamefont{Batten, Stillinger,
  and Torquato}}]{Ba08}
\bibinfo{author}{\bibfnamefont{R.~D.} \bibnamefont{Batten}},
  \bibinfo{author}{\bibfnamefont{F.~H.} \bibnamefont{Stillinger}},
  \bibnamefont{and} \bibinfo{author}{\bibfnamefont{S.}~\bibnamefont{Torquato}},
  \bibinfo{journal}{J. Appl. Phys.} \textbf{\bibinfo{volume}{104}},
  \bibinfo{pages}{033504} (\bibinfo{year}{2008}).

\bibitem[{\citenamefont{Batten et~al.}(2009)\citenamefont{Batten, Stillinger,
  and Torquato}}]{Ba09}
\bibinfo{author}{\bibfnamefont{R.~D.} \bibnamefont{Batten}},
  \bibinfo{author}{\bibfnamefont{F.~H.} \bibnamefont{Stillinger}},
  \bibnamefont{and} \bibinfo{author}{\bibfnamefont{S.}~\bibnamefont{Torquato}},
  \bibinfo{journal}{Phys. Rev. Lett.} \textbf{\bibinfo{volume}{103}},
  \bibinfo{pages}{050602} (\bibinfo{year}{2009}).

\bibitem[{\citenamefont{Jiao}(2022)}]{jiao2022hyperuniformity}
\bibinfo{author}{\bibfnamefont{Y.}~\bibnamefont{Jiao}},
  \bibinfo{journal}{Physica A: Statistical Mechanics and its Applications}
  \textbf{\bibinfo{volume}{585}}, \bibinfo{pages}{126435}
  (\bibinfo{year}{2022}).

\bibitem[{\citenamefont{Torquato et~al.}(2008)\citenamefont{Torquato,
  Scardicchio, and Zachary}}]{To08}
\bibinfo{author}{\bibfnamefont{S.}~\bibnamefont{Torquato}},
  \bibinfo{author}{\bibfnamefont{A.}~\bibnamefont{Scardicchio}},
  \bibnamefont{and} \bibinfo{author}{\bibfnamefont{C.~E.}
  \bibnamefont{Zachary}}, \bibinfo{journal}{J. Stat. Mech.: Theory Exp.} p.
  \bibinfo{pages}{P11019} (\bibinfo{year}{2008}).

\bibitem[{\citenamefont{Feynman and Cohen}(1956)}]{Fe56}
\bibinfo{author}{\bibfnamefont{R.~P.} \bibnamefont{Feynman}} \bibnamefont{and}
  \bibinfo{author}{\bibfnamefont{M.}~\bibnamefont{Cohen}},
  \bibinfo{journal}{Phys. Rev.} \textbf{\bibinfo{volume}{102}},
  \bibinfo{pages}{1189} (\bibinfo{year}{1956}).

\bibitem[{\citenamefont{Kurita and Weeks}(2011)}]{Ku11}
\bibinfo{author}{\bibfnamefont{R.}~\bibnamefont{Kurita}} \bibnamefont{and}
  \bibinfo{author}{\bibfnamefont{E.~R.} \bibnamefont{Weeks}},
  \bibinfo{journal}{Phys. Rev. E} \textbf{\bibinfo{volume}{84}},
  \bibinfo{pages}{030401} (\bibinfo{year}{2011}).

\bibitem[{\citenamefont{Hunter and Weeks}(2012)}]{Hu12}
\bibinfo{author}{\bibfnamefont{G.~L.} \bibnamefont{Hunter}} \bibnamefont{and}
  \bibinfo{author}{\bibfnamefont{E.~R.} \bibnamefont{Weeks}},
  \bibinfo{journal}{Rep. Prog. Phys.} \textbf{\bibinfo{volume}{75}},
  \bibinfo{pages}{066501} (\bibinfo{year}{2012}).

\bibitem[{\citenamefont{Dreyfus et~al.}(2015)\citenamefont{Dreyfus, Xu, Still,
  Hough, Yodh, and Torquato}}]{Dr15}
\bibinfo{author}{\bibfnamefont{R.}~\bibnamefont{Dreyfus}},
  \bibinfo{author}{\bibfnamefont{Y.}~\bibnamefont{Xu}},
  \bibinfo{author}{\bibfnamefont{T.}~\bibnamefont{Still}},
  \bibinfo{author}{\bibfnamefont{L.~A.} \bibnamefont{Hough}},
  \bibinfo{author}{\bibfnamefont{A.~G.} \bibnamefont{Yodh}}, \bibnamefont{and}
  \bibinfo{author}{\bibfnamefont{S.}~\bibnamefont{Torquato}},
  \bibinfo{journal}{Phys. Rev. E} \textbf{\bibinfo{volume}{91}},
  \bibinfo{pages}{012302} (\bibinfo{year}{2015}).

\bibitem[{\citenamefont{Hexner and Levine}(2015)}]{He15}
\bibinfo{author}{\bibfnamefont{D.}~\bibnamefont{Hexner}} \bibnamefont{and}
  \bibinfo{author}{\bibfnamefont{D.}~\bibnamefont{Levine}},
  \bibinfo{journal}{Phys. Rev. Lett.} \textbf{\bibinfo{volume}{114}},
  \bibinfo{pages}{110602} (\bibinfo{year}{2015}).

\bibitem[{\citenamefont{Jack et~al.}(2015)\citenamefont{Jack, Thompson, and
  Sollich}}]{Ja15}
\bibinfo{author}{\bibfnamefont{R.~L.} \bibnamefont{Jack}},
  \bibinfo{author}{\bibfnamefont{I.~R.} \bibnamefont{Thompson}},
  \bibnamefont{and} \bibinfo{author}{\bibfnamefont{P.}~\bibnamefont{Sollich}},
  \bibinfo{journal}{Phys. Rev. Lett.} \textbf{\bibinfo{volume}{114}},
  \bibinfo{pages}{060601} (\bibinfo{year}{2015}).

\bibitem[{\citenamefont{Weijs et~al.}(2015)\citenamefont{Weijs, Jeanneret,
  Dreyfus, and Bartolo}}]{We15}
\bibinfo{author}{\bibfnamefont{J.~H.} \bibnamefont{Weijs}},
  \bibinfo{author}{\bibfnamefont{R.}~\bibnamefont{Jeanneret}},
  \bibinfo{author}{\bibfnamefont{R.}~\bibnamefont{Dreyfus}}, \bibnamefont{and}
  \bibinfo{author}{\bibfnamefont{D.}~\bibnamefont{Bartolo}},
  \bibinfo{journal}{Phys. Rev. Lett.} \textbf{\bibinfo{volume}{115}},
  \bibinfo{pages}{108301} (\bibinfo{year}{2015}).

\bibitem[{\citenamefont{Salvalaglio et~al.}(2020)\citenamefont{Salvalaglio,
  Bouabdellaoui, Bollani, Benali, Favre, Claude, Wenger, de~Anna, Intonti,
  Voigt et~al.}}]{salvalaglio2020hyperuniform}
\bibinfo{author}{\bibfnamefont{M.}~\bibnamefont{Salvalaglio}},
  \bibinfo{author}{\bibfnamefont{M.}~\bibnamefont{Bouabdellaoui}},
  \bibinfo{author}{\bibfnamefont{M.}~\bibnamefont{Bollani}},
  \bibinfo{author}{\bibfnamefont{A.}~\bibnamefont{Benali}},
  \bibinfo{author}{\bibfnamefont{L.}~\bibnamefont{Favre}},
  \bibinfo{author}{\bibfnamefont{J.-B.} \bibnamefont{Claude}},
  \bibinfo{author}{\bibfnamefont{J.}~\bibnamefont{Wenger}},
  \bibinfo{author}{\bibfnamefont{P.}~\bibnamefont{de~Anna}},
  \bibinfo{author}{\bibfnamefont{F.}~\bibnamefont{Intonti}},
  \bibinfo{author}{\bibfnamefont{A.}~\bibnamefont{Voigt}},
  \bibnamefont{et~al.}, \bibinfo{journal}{Physical Review Letters}
  \textbf{\bibinfo{volume}{125}}, \bibinfo{pages}{126101}
  (\bibinfo{year}{2020}).

\bibitem[{\citenamefont{Nizam et~al.}(2021)\citenamefont{Nizam, Makey, Barbier,
  Kahraman, Demir, Shafigh, Galioglu, Vahabli, H{\"u}sn{\"u}gil,
  G{\"u}ne{\c{s}} et~al.}}]{nizam2021dynamic}
\bibinfo{author}{\bibfnamefont{{\"U}.~S.} \bibnamefont{Nizam}},
  \bibinfo{author}{\bibfnamefont{G.}~\bibnamefont{Makey}},
  \bibinfo{author}{\bibfnamefont{M.}~\bibnamefont{Barbier}},
  \bibinfo{author}{\bibfnamefont{S.~S.} \bibnamefont{Kahraman}},
  \bibinfo{author}{\bibfnamefont{E.}~\bibnamefont{Demir}},
  \bibinfo{author}{\bibfnamefont{E.~E.} \bibnamefont{Shafigh}},
  \bibinfo{author}{\bibfnamefont{S.}~\bibnamefont{Galioglu}},
  \bibinfo{author}{\bibfnamefont{D.}~\bibnamefont{Vahabli}},
  \bibinfo{author}{\bibfnamefont{S.}~\bibnamefont{H{\"u}sn{\"u}gil}},
  \bibinfo{author}{\bibfnamefont{M.~H.} \bibnamefont{G{\"u}ne{\c{s}}}},
  \bibnamefont{et~al.}, \bibinfo{journal}{Journal of Physics: Condensed Matter}
  \textbf{\bibinfo{volume}{33}}, \bibinfo{pages}{304002}
  (\bibinfo{year}{2021}).

\bibitem[{\citenamefont{Zheng et~al.}(2023)\citenamefont{Zheng, Klatt, and
  L{\"o}wen}}]{zheng2023universal}
\bibinfo{author}{\bibfnamefont{Y.}~\bibnamefont{Zheng}},
  \bibinfo{author}{\bibfnamefont{M.~A.} \bibnamefont{Klatt}}, \bibnamefont{and}
  \bibinfo{author}{\bibfnamefont{H.}~\bibnamefont{L{\"o}wen}},
  \bibinfo{journal}{arXiv preprint arXiv:2310.03107}  (\bibinfo{year}{2023}).

\bibitem[{\citenamefont{Lei et~al.}(2019)\citenamefont{Lei, Ciamarra, and
  Ni}}]{Le19}
\bibinfo{author}{\bibfnamefont{Q.-L.} \bibnamefont{Lei}},
  \bibinfo{author}{\bibfnamefont{M.~P.} \bibnamefont{Ciamarra}},
  \bibnamefont{and} \bibinfo{author}{\bibfnamefont{R.}~\bibnamefont{Ni}},
  \bibinfo{journal}{Sci. Adv.} \textbf{\bibinfo{volume}{5}},
  \bibinfo{pages}{eaau7423} (\bibinfo{year}{2019}).

\bibitem[{\citenamefont{Lei and Ni}(2019)}]{lei2019hydrodynamics}
\bibinfo{author}{\bibfnamefont{Q.-L.} \bibnamefont{Lei}} \bibnamefont{and}
  \bibinfo{author}{\bibfnamefont{R.}~\bibnamefont{Ni}},
  \bibinfo{journal}{Proceedings of the National Academy of Sciences}
  \textbf{\bibinfo{volume}{116}}, \bibinfo{pages}{22983}
  (\bibinfo{year}{2019}).

\bibitem[{\citenamefont{Huang et~al.}(2021)\citenamefont{Huang, Hu, Yang, Liu,
  and Zhang}}]{huang2021circular}
\bibinfo{author}{\bibfnamefont{M.}~\bibnamefont{Huang}},
  \bibinfo{author}{\bibfnamefont{W.}~\bibnamefont{Hu}},
  \bibinfo{author}{\bibfnamefont{S.}~\bibnamefont{Yang}},
  \bibinfo{author}{\bibfnamefont{Q.-X.} \bibnamefont{Liu}}, \bibnamefont{and}
  \bibinfo{author}{\bibfnamefont{H.}~\bibnamefont{Zhang}},
  \bibinfo{journal}{Proceedings of the National Academy of Sciences}
  \textbf{\bibinfo{volume}{118}}, \bibinfo{pages}{e2100493118}
  (\bibinfo{year}{2021}).

\bibitem[{\citenamefont{Zhang and Snezhko}(2022)}]{zhang2022hyperuniform}
\bibinfo{author}{\bibfnamefont{B.}~\bibnamefont{Zhang}} \bibnamefont{and}
  \bibinfo{author}{\bibfnamefont{A.}~\bibnamefont{Snezhko}},
  \bibinfo{journal}{Physical Review Letters} \textbf{\bibinfo{volume}{128}},
  \bibinfo{pages}{218002} (\bibinfo{year}{2022}).

\bibitem[{\citenamefont{Oppenheimer et~al.}(2022)\citenamefont{Oppenheimer,
  Stein, Zion, and Shelley}}]{oppenheimer2022hyperuniformity}
\bibinfo{author}{\bibfnamefont{N.}~\bibnamefont{Oppenheimer}},
  \bibinfo{author}{\bibfnamefont{D.~B.} \bibnamefont{Stein}},
  \bibinfo{author}{\bibfnamefont{M.~Y.~B.} \bibnamefont{Zion}},
  \bibnamefont{and} \bibinfo{author}{\bibfnamefont{M.~J.}
  \bibnamefont{Shelley}}, \bibinfo{journal}{Nature communications}
  \textbf{\bibinfo{volume}{13}}, \bibinfo{pages}{804} (\bibinfo{year}{2022}).

\bibitem[{\citenamefont{Hexner and Levine}(2017)}]{hexner2017noise}
\bibinfo{author}{\bibfnamefont{D.}~\bibnamefont{Hexner}} \bibnamefont{and}
  \bibinfo{author}{\bibfnamefont{D.}~\bibnamefont{Levine}},
  \bibinfo{journal}{Physical review letters} \textbf{\bibinfo{volume}{118}},
  \bibinfo{pages}{020601} (\bibinfo{year}{2017}).

\bibitem[{\citenamefont{Hexner et~al.}(2017)\citenamefont{Hexner, Chaikin, and
  Levine}}]{hexner2017enhanced}
\bibinfo{author}{\bibfnamefont{D.}~\bibnamefont{Hexner}},
  \bibinfo{author}{\bibfnamefont{P.~M.} \bibnamefont{Chaikin}},
  \bibnamefont{and} \bibinfo{author}{\bibfnamefont{D.}~\bibnamefont{Levine}},
  \bibinfo{journal}{Proceedings of the National Academy of Sciences}
  \textbf{\bibinfo{volume}{114}}, \bibinfo{pages}{4294} (\bibinfo{year}{2017}).

\bibitem[{\citenamefont{Weijs and Bartolo}(2017)}]{weijs2017mixing}
\bibinfo{author}{\bibfnamefont{J.~H.} \bibnamefont{Weijs}} \bibnamefont{and}
  \bibinfo{author}{\bibfnamefont{D.}~\bibnamefont{Bartolo}},
  \bibinfo{journal}{Physical review letters} \textbf{\bibinfo{volume}{119}},
  \bibinfo{pages}{048002} (\bibinfo{year}{2017}).

\bibitem[{\citenamefont{Wilken et~al.}(2022)\citenamefont{Wilken, Guo, Levine,
  and Chaikin}}]{wilken2022random}
\bibinfo{author}{\bibfnamefont{S.}~\bibnamefont{Wilken}},
  \bibinfo{author}{\bibfnamefont{A.~Z.} \bibnamefont{Guo}},
  \bibinfo{author}{\bibfnamefont{D.}~\bibnamefont{Levine}}, \bibnamefont{and}
  \bibinfo{author}{\bibfnamefont{P.~M.} \bibnamefont{Chaikin}},
  \bibinfo{journal}{arXiv preprint arXiv:2212.09913}  (\bibinfo{year}{2022}).

\bibitem[{\citenamefont{Jiao et~al.}(2014)\citenamefont{Jiao, Lau, Hatzikirou,
  Meyer-Hermann, Corbo, and Torquato}}]{Ji14}
\bibinfo{author}{\bibfnamefont{Y.}~\bibnamefont{Jiao}},
  \bibinfo{author}{\bibfnamefont{T.}~\bibnamefont{Lau}},
  \bibinfo{author}{\bibfnamefont{H.}~\bibnamefont{Hatzikirou}},
  \bibinfo{author}{\bibfnamefont{M.}~\bibnamefont{Meyer-Hermann}},
  \bibinfo{author}{\bibfnamefont{J.~C.} \bibnamefont{Corbo}}, \bibnamefont{and}
  \bibinfo{author}{\bibfnamefont{S.}~\bibnamefont{Torquato}},
  \bibinfo{journal}{Phys. Rev. E} \textbf{\bibinfo{volume}{89}},
  \bibinfo{pages}{022721} (\bibinfo{year}{2014}).

\bibitem[{\citenamefont{Mayer et~al.}(2015)\citenamefont{Mayer,
  Balasubramanian, Mora, and Walczak}}]{Ma15}
\bibinfo{author}{\bibfnamefont{A.}~\bibnamefont{Mayer}},
  \bibinfo{author}{\bibfnamefont{V.}~\bibnamefont{Balasubramanian}},
  \bibinfo{author}{\bibfnamefont{T.}~\bibnamefont{Mora}}, \bibnamefont{and}
  \bibinfo{author}{\bibfnamefont{A.~M.} \bibnamefont{Walczak}},
  \bibinfo{journal}{Proc. Natl. Acad. Sci. USA} \textbf{\bibinfo{volume}{112}},
  \bibinfo{pages}{5950} (\bibinfo{year}{2015}).

\bibitem[{\citenamefont{Ge}(2023)}]{ge2023hidden}
\bibinfo{author}{\bibfnamefont{Z.}~\bibnamefont{Ge}},
  \bibinfo{journal}{Proceedings of the National Academy of Sciences}
  \textbf{\bibinfo{volume}{120}}, \bibinfo{pages}{e2306514120}
  (\bibinfo{year}{2023}).

\bibitem[{\citenamefont{Rumi et~al.}(2019)\citenamefont{Rumi, S{\'a}nchez,
  El{\'\i}as, Maldonado, Puig, Bolecek, Nieva, Konczykowski, Fasano, and
  Kolton}}]{Ru19}
\bibinfo{author}{\bibfnamefont{G.}~\bibnamefont{Rumi}},
  \bibinfo{author}{\bibfnamefont{J.~A.} \bibnamefont{S{\'a}nchez}},
  \bibinfo{author}{\bibfnamefont{F.}~\bibnamefont{El{\'\i}as}},
  \bibinfo{author}{\bibfnamefont{R.~C.} \bibnamefont{Maldonado}},
  \bibinfo{author}{\bibfnamefont{J.}~\bibnamefont{Puig}},
  \bibinfo{author}{\bibfnamefont{N.~R.~C.} \bibnamefont{Bolecek}},
  \bibinfo{author}{\bibfnamefont{G.}~\bibnamefont{Nieva}},
  \bibinfo{author}{\bibfnamefont{M.}~\bibnamefont{Konczykowski}},
  \bibinfo{author}{\bibfnamefont{Y.}~\bibnamefont{Fasano}}, \bibnamefont{and}
  \bibinfo{author}{\bibfnamefont{A.~B.} \bibnamefont{Kolton}},
  \bibinfo{journal}{Phys. Rev. Res.} \textbf{\bibinfo{volume}{1}},
  \bibinfo{pages}{033057} (\bibinfo{year}{2019}).

\bibitem[{\citenamefont{S{\'a}nchez et~al.}(2019)\citenamefont{S{\'a}nchez,
  Maldonado, Bolecek, Rumi, Pedrazzini, Dolz, Nieva, van~der Beek,
  Konczykowski, Dewhurst et~al.}}]{Sa19}
\bibinfo{author}{\bibfnamefont{J.~A.} \bibnamefont{S{\'a}nchez}},
  \bibinfo{author}{\bibfnamefont{R.~C.} \bibnamefont{Maldonado}},
  \bibinfo{author}{\bibfnamefont{N.~R.~C.} \bibnamefont{Bolecek}},
  \bibinfo{author}{\bibfnamefont{G.}~\bibnamefont{Rumi}},
  \bibinfo{author}{\bibfnamefont{P.}~\bibnamefont{Pedrazzini}},
  \bibinfo{author}{\bibfnamefont{M.~I.} \bibnamefont{Dolz}},
  \bibinfo{author}{\bibfnamefont{G.}~\bibnamefont{Nieva}},
  \bibinfo{author}{\bibfnamefont{C.~J.} \bibnamefont{van~der Beek}},
  \bibinfo{author}{\bibfnamefont{M.}~\bibnamefont{Konczykowski}},
  \bibinfo{author}{\bibfnamefont{C.~D.} \bibnamefont{Dewhurst}},
  \bibnamefont{et~al.}, \bibinfo{journal}{Commun. Phys.}
  \textbf{\bibinfo{volume}{2}}, \bibinfo{pages}{1} (\bibinfo{year}{2019}).

\bibitem[{\citenamefont{Gerasimenko et~al.}(2019)\citenamefont{Gerasimenko,
  Vaskivskyi, Litskevich, Ravnik, Vodeb, Diego, Kabanov, and
  Mihailovic}}]{Ge19quantum}
\bibinfo{author}{\bibfnamefont{Y.~A.} \bibnamefont{Gerasimenko}},
  \bibinfo{author}{\bibfnamefont{I.}~\bibnamefont{Vaskivskyi}},
  \bibinfo{author}{\bibfnamefont{M.}~\bibnamefont{Litskevich}},
  \bibinfo{author}{\bibfnamefont{J.}~\bibnamefont{Ravnik}},
  \bibinfo{author}{\bibfnamefont{J.}~\bibnamefont{Vodeb}},
  \bibinfo{author}{\bibfnamefont{M.}~\bibnamefont{Diego}},
  \bibinfo{author}{\bibfnamefont{V.}~\bibnamefont{Kabanov}}, \bibnamefont{and}
  \bibinfo{author}{\bibfnamefont{D.}~\bibnamefont{Mihailovic}},
  \bibinfo{journal}{Nat. Mater.} \textbf{\bibinfo{volume}{18}},
  \bibinfo{pages}{1078} (\bibinfo{year}{2019}).

\bibitem[{\citenamefont{Sakai et~al.}(2022)\citenamefont{Sakai, Arita, and
  Ohtsuki}}]{sakai2022quantum}
\bibinfo{author}{\bibfnamefont{S.}~\bibnamefont{Sakai}},
  \bibinfo{author}{\bibfnamefont{R.}~\bibnamefont{Arita}}, \bibnamefont{and}
  \bibinfo{author}{\bibfnamefont{T.}~\bibnamefont{Ohtsuki}},
  \bibinfo{journal}{arXiv preprint arXiv:2207.09698}  (\bibinfo{year}{2022}).

\bibitem[{\citenamefont{Chen et~al.}(2023{\natexlab{a}})\citenamefont{Chen,
  Zhuang, Chen, Huang, Vlcek, and Jiao}}]{chen2023disordered}
\bibinfo{author}{\bibfnamefont{D.}~\bibnamefont{Chen}},
  \bibinfo{author}{\bibfnamefont{H.}~\bibnamefont{Zhuang}},
  \bibinfo{author}{\bibfnamefont{M.}~\bibnamefont{Chen}},
  \bibinfo{author}{\bibfnamefont{P.~Y.} \bibnamefont{Huang}},
  \bibinfo{author}{\bibfnamefont{V.}~\bibnamefont{Vlcek}}, \bibnamefont{and}
  \bibinfo{author}{\bibfnamefont{Y.}~\bibnamefont{Jiao}},
  \bibinfo{journal}{Applied Physics Reviews} \textbf{\bibinfo{volume}{10}}
  (\bibinfo{year}{2023}{\natexlab{a}}).

\bibitem[{\citenamefont{Zheng et~al.}(2020)\citenamefont{Zheng, Liu, Nan, Shen,
  Zhang, Chen, He, Xu, Chen, Jiao et~al.}}]{Zh20}
\bibinfo{author}{\bibfnamefont{Y.}~\bibnamefont{Zheng}},
  \bibinfo{author}{\bibfnamefont{L.}~\bibnamefont{Liu}},
  \bibinfo{author}{\bibfnamefont{H.}~\bibnamefont{Nan}},
  \bibinfo{author}{\bibfnamefont{Z.-X.} \bibnamefont{Shen}},
  \bibinfo{author}{\bibfnamefont{G.}~\bibnamefont{Zhang}},
  \bibinfo{author}{\bibfnamefont{D.}~\bibnamefont{Chen}},
  \bibinfo{author}{\bibfnamefont{L.}~\bibnamefont{He}},
  \bibinfo{author}{\bibfnamefont{W.}~\bibnamefont{Xu}},
  \bibinfo{author}{\bibfnamefont{M.}~\bibnamefont{Chen}},
  \bibinfo{author}{\bibfnamefont{Y.}~\bibnamefont{Jiao}}, \bibnamefont{et~al.},
  \bibinfo{journal}{Sci. Adv.} \textbf{\bibinfo{volume}{6}},
  \bibinfo{pages}{eaba0826} (\bibinfo{year}{2020}).

\bibitem[{\citenamefont{Chen et~al.}(2021{\natexlab{a}})\citenamefont{Chen,
  Zheng, Liu, Zhang, Chen, Jiao, and Zhuang}}]{Ch21}
\bibinfo{author}{\bibfnamefont{D.}~\bibnamefont{Chen}},
  \bibinfo{author}{\bibfnamefont{Y.}~\bibnamefont{Zheng}},
  \bibinfo{author}{\bibfnamefont{L.}~\bibnamefont{Liu}},
  \bibinfo{author}{\bibfnamefont{G.}~\bibnamefont{Zhang}},
  \bibinfo{author}{\bibfnamefont{M.}~\bibnamefont{Chen}},
  \bibinfo{author}{\bibfnamefont{Y.}~\bibnamefont{Jiao}}, \bibnamefont{and}
  \bibinfo{author}{\bibfnamefont{H.}~\bibnamefont{Zhuang}},
  \bibinfo{journal}{Proc. Natl. Acad. Sci. U.S.A.}
  \textbf{\bibinfo{volume}{118}}, \bibinfo{pages}{e2016862118}
  (\bibinfo{year}{2021}{\natexlab{a}}).

\bibitem[{\citenamefont{Chen et~al.}(2021{\natexlab{b}})\citenamefont{Chen,
  Zheng, Lee, Kang, Zhu, Zhuang, Huang, and Jiao}}]{PhysRevB.103.224102}
\bibinfo{author}{\bibfnamefont{D.}~\bibnamefont{Chen}},
  \bibinfo{author}{\bibfnamefont{Y.}~\bibnamefont{Zheng}},
  \bibinfo{author}{\bibfnamefont{C.-H.} \bibnamefont{Lee}},
  \bibinfo{author}{\bibfnamefont{S.}~\bibnamefont{Kang}},
  \bibinfo{author}{\bibfnamefont{W.}~\bibnamefont{Zhu}},
  \bibinfo{author}{\bibfnamefont{H.}~\bibnamefont{Zhuang}},
  \bibinfo{author}{\bibfnamefont{P.~Y.} \bibnamefont{Huang}}, \bibnamefont{and}
  \bibinfo{author}{\bibfnamefont{Y.}~\bibnamefont{Jiao}},
  \bibinfo{journal}{Phys. Rev. B} \textbf{\bibinfo{volume}{103}},
  \bibinfo{pages}{224102} (\bibinfo{year}{2021}{\natexlab{b}}),
  \urlprefix\url{https://link.aps.org/doi/10.1103/PhysRevB.103.224102}.

\bibitem[{\citenamefont{Zheng et~al.}(2021)\citenamefont{Zheng, Chen, Liu, Liu,
  Chen, Zhuang, and Jiao}}]{Zh21}
\bibinfo{author}{\bibfnamefont{Y.}~\bibnamefont{Zheng}},
  \bibinfo{author}{\bibfnamefont{D.}~\bibnamefont{Chen}},
  \bibinfo{author}{\bibfnamefont{L.}~\bibnamefont{Liu}},
  \bibinfo{author}{\bibfnamefont{Y.}~\bibnamefont{Liu}},
  \bibinfo{author}{\bibfnamefont{M.}~\bibnamefont{Chen}},
  \bibinfo{author}{\bibfnamefont{H.}~\bibnamefont{Zhuang}}, \bibnamefont{and}
  \bibinfo{author}{\bibfnamefont{Y.}~\bibnamefont{Jiao}},
  \bibinfo{journal}{Phys. Rev. B} \textbf{\bibinfo{volume}{103}},
  \bibinfo{pages}{245413} (\bibinfo{year}{2021}).

\bibitem[{\citenamefont{Chen et~al.}(2023{\natexlab{b}})\citenamefont{Chen,
  Jiang, Wang, Zhuang, and Jiao}}]{chen2021multihyperuniform}
\bibinfo{author}{\bibfnamefont{D.}~\bibnamefont{Chen}},
  \bibinfo{author}{\bibfnamefont{X.}~\bibnamefont{Jiang}},
  \bibinfo{author}{\bibfnamefont{D.}~\bibnamefont{Wang}},
  \bibinfo{author}{\bibfnamefont{H.}~\bibnamefont{Zhuang}}, \bibnamefont{and}
  \bibinfo{author}{\bibfnamefont{Y.}~\bibnamefont{Jiao}},
  \bibinfo{journal}{Acta Materialia} \textbf{\bibinfo{volume}{246}},
  \bibinfo{pages}{118678} (\bibinfo{year}{2023}{\natexlab{b}}).

\bibitem[{\citenamefont{Chen et~al.}(2022)\citenamefont{Chen, Liu, Zhuang,
  Chen, and Jiao}}]{nanotube}
\bibinfo{author}{\bibfnamefont{D.}~\bibnamefont{Chen}},
  \bibinfo{author}{\bibfnamefont{Y.}~\bibnamefont{Liu}},
  \bibinfo{author}{\bibfnamefont{H.}~\bibnamefont{Zhuang}},
  \bibinfo{author}{\bibfnamefont{M.}~\bibnamefont{Chen}}, \bibnamefont{and}
  \bibinfo{author}{\bibfnamefont{Y.}~\bibnamefont{Jiao}},
  \bibinfo{journal}{Physical Review B} \textbf{\bibinfo{volume}{106}},
  \bibinfo{pages}{235427} (\bibinfo{year}{2022}).

\bibitem[{\citenamefont{Zhang et~al.}(2023)\citenamefont{Zhang, Wang, Zhang,
  Yu, and Douglas}}]{zhang2023approach}
\bibinfo{author}{\bibfnamefont{H.}~\bibnamefont{Zhang}},
  \bibinfo{author}{\bibfnamefont{X.}~\bibnamefont{Wang}},
  \bibinfo{author}{\bibfnamefont{J.}~\bibnamefont{Zhang}},
  \bibinfo{author}{\bibfnamefont{H.-B.} \bibnamefont{Yu}}, \bibnamefont{and}
  \bibinfo{author}{\bibfnamefont{J.~F.} \bibnamefont{Douglas}},
  \bibinfo{journal}{arXiv preprint arXiv:2302.01429}  (\bibinfo{year}{2023}).

\bibitem[{\citenamefont{Chremos and Douglas}(2018)}]{Ch18b}
\bibinfo{author}{\bibfnamefont{A.}~\bibnamefont{Chremos}} \bibnamefont{and}
  \bibinfo{author}{\bibfnamefont{J.~F.} \bibnamefont{Douglas}},
  \bibinfo{journal}{Phys. Rev. Lett.} \textbf{\bibinfo{volume}{121}},
  \bibinfo{pages}{258002} (\bibinfo{year}{2018}).

\bibitem[{\citenamefont{Florescu et~al.}(2009)\citenamefont{Florescu, Torquato,
  and Steinhardt}}]{Fl09}
\bibinfo{author}{\bibfnamefont{M.}~\bibnamefont{Florescu}},
  \bibinfo{author}{\bibfnamefont{S.}~\bibnamefont{Torquato}}, \bibnamefont{and}
  \bibinfo{author}{\bibfnamefont{P.~J.} \bibnamefont{Steinhardt}},
  \bibinfo{journal}{Proc. Natl. Acad. Sci. U.S.A.}
  \textbf{\bibinfo{volume}{106}}, \bibinfo{pages}{20658}
  (\bibinfo{year}{2009}).

\bibitem[{\citenamefont{Man et~al.}(2013)\citenamefont{Man, Florescu,
  Williamson, He, Hashemizad, Leung, Liner, Torquato, Chaikin, and
  Steinhardt}}]{Ma13}
\bibinfo{author}{\bibfnamefont{W.}~\bibnamefont{Man}},
  \bibinfo{author}{\bibfnamefont{M.}~\bibnamefont{Florescu}},
  \bibinfo{author}{\bibfnamefont{E.~P.} \bibnamefont{Williamson}},
  \bibinfo{author}{\bibfnamefont{Y.}~\bibnamefont{He}},
  \bibinfo{author}{\bibfnamefont{S.~R.} \bibnamefont{Hashemizad}},
  \bibinfo{author}{\bibfnamefont{B.~Y.~C.} \bibnamefont{Leung}},
  \bibinfo{author}{\bibfnamefont{D.~R.} \bibnamefont{Liner}},
  \bibinfo{author}{\bibfnamefont{S.}~\bibnamefont{Torquato}},
  \bibinfo{author}{\bibfnamefont{P.~M.} \bibnamefont{Chaikin}},
  \bibnamefont{and} \bibinfo{author}{\bibfnamefont{P.~J.}
  \bibnamefont{Steinhardt}}, \bibinfo{journal}{Proc. Natl. Acad. Sci. U.S.A.}
  \textbf{\bibinfo{volume}{110}}, \bibinfo{pages}{15886}
  (\bibinfo{year}{2013}).

\bibitem[{\citenamefont{Klatt et~al.}(2022)\citenamefont{Klatt, Steinhardt, and
  Torquato}}]{klatt2022wave}
\bibinfo{author}{\bibfnamefont{M.~A.} \bibnamefont{Klatt}},
  \bibinfo{author}{\bibfnamefont{P.~J.} \bibnamefont{Steinhardt}},
  \bibnamefont{and} \bibinfo{author}{\bibfnamefont{S.}~\bibnamefont{Torquato}},
  \bibinfo{journal}{Proceedings of the National Academy of Sciences}
  \textbf{\bibinfo{volume}{119}}, \bibinfo{pages}{e2213633119}
  (\bibinfo{year}{2022}).

\bibitem[{\citenamefont{Zhang et~al.}(2016)\citenamefont{Zhang, Stillinger, and
  Torquato}}]{Zh16}
\bibinfo{author}{\bibfnamefont{G.}~\bibnamefont{Zhang}},
  \bibinfo{author}{\bibfnamefont{F.~H.} \bibnamefont{Stillinger}},
  \bibnamefont{and} \bibinfo{author}{\bibfnamefont{S.}~\bibnamefont{Torquato}},
  \bibinfo{journal}{J. Chem. Phys.} \textbf{\bibinfo{volume}{145}},
  \bibinfo{pages}{244109} (\bibinfo{year}{2016}).

\bibitem[{\citenamefont{Chen and Torquato}(2018)}]{Ch18a}
\bibinfo{author}{\bibfnamefont{D.}~\bibnamefont{Chen}} \bibnamefont{and}
  \bibinfo{author}{\bibfnamefont{S.}~\bibnamefont{Torquato}},
  \bibinfo{journal}{Acta Mater.} \textbf{\bibinfo{volume}{142}},
  \bibinfo{pages}{152} (\bibinfo{year}{2018}).

\bibitem[{\citenamefont{Torquato}(2021)}]{torquato2021diffusion}
\bibinfo{author}{\bibfnamefont{S.}~\bibnamefont{Torquato}},
  \bibinfo{journal}{Physical Review E} \textbf{\bibinfo{volume}{104}},
  \bibinfo{pages}{054102} (\bibinfo{year}{2021}).

\bibitem[{\citenamefont{Xu et~al.}(2017)\citenamefont{Xu, Chen, Chen, Xu, and
  Jiao}}]{Xu17}
\bibinfo{author}{\bibfnamefont{Y.}~\bibnamefont{Xu}},
  \bibinfo{author}{\bibfnamefont{S.}~\bibnamefont{Chen}},
  \bibinfo{author}{\bibfnamefont{P.}~\bibnamefont{Chen}},
  \bibinfo{author}{\bibfnamefont{W.}~\bibnamefont{Xu}}, \bibnamefont{and}
  \bibinfo{author}{\bibfnamefont{Y.}~\bibnamefont{Jiao}},
  \bibinfo{journal}{Phys. Rev. E} \textbf{\bibinfo{volume}{96}},
  \bibinfo{pages}{043301} (\bibinfo{year}{2017}).

\bibitem[{\citenamefont{Klatt and Torquato}(2018)}]{Kl18}
\bibinfo{author}{\bibfnamefont{M.~A.} \bibnamefont{Klatt}} \bibnamefont{and}
  \bibinfo{author}{\bibfnamefont{S.}~\bibnamefont{Torquato}},
  \bibinfo{journal}{Phys. Rev. E} \textbf{\bibinfo{volume}{97}},
  \bibinfo{pages}{012118} (\bibinfo{year}{2018}).

\bibitem[{\citenamefont{Leseur et~al.}(2016)\citenamefont{Leseur, Pierrat, and
  Carminati}}]{Le16}
\bibinfo{author}{\bibfnamefont{O.}~\bibnamefont{Leseur}},
  \bibinfo{author}{\bibfnamefont{R.}~\bibnamefont{Pierrat}}, \bibnamefont{and}
  \bibinfo{author}{\bibfnamefont{R.}~\bibnamefont{Carminati}},
  \bibinfo{journal}{Optica} \textbf{\bibinfo{volume}{3}}, \bibinfo{pages}{763}
  (\bibinfo{year}{2016}).

\bibitem[{\citenamefont{Kim and Torquato}(2023)}]{kim2023effective}
\bibinfo{author}{\bibfnamefont{J.}~\bibnamefont{Kim}} \bibnamefont{and}
  \bibinfo{author}{\bibfnamefont{S.}~\bibnamefont{Torquato}},
  \bibinfo{journal}{arXiv preprint arXiv:2305.13280}  (\bibinfo{year}{2023}).

\bibitem[{\citenamefont{Yu}(2023)}]{yu2023evolving}
\bibinfo{author}{\bibfnamefont{S.}~\bibnamefont{Yu}}, \bibinfo{journal}{Nature
  Computational Science} \textbf{\bibinfo{volume}{3}}, \bibinfo{pages}{128}
  (\bibinfo{year}{2023}).

\bibitem[{\citenamefont{Shi et~al.}(2023)\citenamefont{Shi, Keeney, Chen, Jiao,
  and Torquato}}]{shi2023computational}
\bibinfo{author}{\bibfnamefont{W.}~\bibnamefont{Shi}},
  \bibinfo{author}{\bibfnamefont{D.}~\bibnamefont{Keeney}},
  \bibinfo{author}{\bibfnamefont{D.}~\bibnamefont{Chen}},
  \bibinfo{author}{\bibfnamefont{Y.}~\bibnamefont{Jiao}}, \bibnamefont{and}
  \bibinfo{author}{\bibfnamefont{S.}~\bibnamefont{Torquato}},
  \bibinfo{journal}{Physical Review E} \textbf{\bibinfo{volume}{108}},
  \bibinfo{pages}{045306} (\bibinfo{year}{2023}).

\bibitem[{\citenamefont{Torquato and Chen}(2018)}]{To18b}
\bibinfo{author}{\bibfnamefont{S.}~\bibnamefont{Torquato}} \bibnamefont{and}
  \bibinfo{author}{\bibfnamefont{D.}~\bibnamefont{Chen}},
  \bibinfo{journal}{Multifunct. Mater.} \textbf{\bibinfo{volume}{1}},
  \bibinfo{pages}{015001} (\bibinfo{year}{2018}).

\bibitem[{\citenamefont{Anderson}(1958)}]{anderson1958absence}
\bibinfo{author}{\bibfnamefont{P.~W.} \bibnamefont{Anderson}},
  \bibinfo{journal}{Physical review} \textbf{\bibinfo{volume}{109}},
  \bibinfo{pages}{1492} (\bibinfo{year}{1958}).

\bibitem[{\citenamefont{Czycholl and Kramer}(1980)}]{czycholl1980numerical}
\bibinfo{author}{\bibfnamefont{G.}~\bibnamefont{Czycholl}} \bibnamefont{and}
  \bibinfo{author}{\bibfnamefont{B.}~\bibnamefont{Kramer}},
  \bibinfo{journal}{Zeitschrift f{\"u}r Physik B Condensed Matter}
  \textbf{\bibinfo{volume}{39}}, \bibinfo{pages}{193} (\bibinfo{year}{1980}).

\bibitem[{\citenamefont{Kramer and MacKinnon}(1993)}]{kramer1993localization}
\bibinfo{author}{\bibfnamefont{B.}~\bibnamefont{Kramer}} \bibnamefont{and}
  \bibinfo{author}{\bibfnamefont{A.}~\bibnamefont{MacKinnon}},
  \bibinfo{journal}{Reports on Progress in Physics}
  \textbf{\bibinfo{volume}{56}}, \bibinfo{pages}{1469} (\bibinfo{year}{1993}).

\bibitem[{\citenamefont{Hohenberg and Kohn}(1964)}]{hohenberg1964inhomogeneous}
\bibinfo{author}{\bibfnamefont{P.}~\bibnamefont{Hohenberg}} \bibnamefont{and}
  \bibinfo{author}{\bibfnamefont{W.}~\bibnamefont{Kohn}},
  \bibinfo{journal}{Physical review} \textbf{\bibinfo{volume}{136}},
  \bibinfo{pages}{B864} (\bibinfo{year}{1964}).

\bibitem[{\citenamefont{Kohn and Sham}(1965)}]{kohn1965self}
\bibinfo{author}{\bibfnamefont{W.}~\bibnamefont{Kohn}} \bibnamefont{and}
  \bibinfo{author}{\bibfnamefont{L.~J.} \bibnamefont{Sham}},
  \bibinfo{journal}{Physical review} \textbf{\bibinfo{volume}{140}},
  \bibinfo{pages}{A1133} (\bibinfo{year}{1965}).

\bibitem[{\citenamefont{Hoffmann et~al.}(1991)\citenamefont{Hoffmann, Janiak,
  and Kollmar}}]{hoffmann1991chemical}
\bibinfo{author}{\bibfnamefont{R.}~\bibnamefont{Hoffmann}},
  \bibinfo{author}{\bibfnamefont{C.}~\bibnamefont{Janiak}}, \bibnamefont{and}
  \bibinfo{author}{\bibfnamefont{C.}~\bibnamefont{Kollmar}},
  \bibinfo{journal}{Macromolecules} \textbf{\bibinfo{volume}{24}},
  \bibinfo{pages}{3725} (\bibinfo{year}{1991}).

\bibitem[{\citenamefont{Curl and Smalley}(1988)}]{curl1988probing}
\bibinfo{author}{\bibfnamefont{R.~F.} \bibnamefont{Curl}} \bibnamefont{and}
  \bibinfo{author}{\bibfnamefont{R.~E.} \bibnamefont{Smalley}},
  \bibinfo{journal}{Science} \textbf{\bibinfo{volume}{242}},
  \bibinfo{pages}{1017} (\bibinfo{year}{1988}).

\bibitem[{\citenamefont{Kresse and
  Furthm{\"u}ller}(1996)}]{kresse1996efficiency}
\bibinfo{author}{\bibfnamefont{G.}~\bibnamefont{Kresse}} \bibnamefont{and}
  \bibinfo{author}{\bibfnamefont{J.}~\bibnamefont{Furthm{\"u}ller}},
  \bibinfo{journal}{Computational materials science}
  \textbf{\bibinfo{volume}{6}}, \bibinfo{pages}{15} (\bibinfo{year}{1996}).

\bibitem[{\citenamefont{Kresse and Furthm\"uller}(1996)}]{PhysRevB.54.11169}
\bibinfo{author}{\bibfnamefont{G.}~\bibnamefont{Kresse}} \bibnamefont{and}
  \bibinfo{author}{\bibfnamefont{J.}~\bibnamefont{Furthm\"uller}},
  \bibinfo{journal}{Phys. Rev. B} \textbf{\bibinfo{volume}{54}},
  \bibinfo{pages}{11169} (\bibinfo{year}{1996}),
  \urlprefix\url{https://link.aps.org/doi/10.1103/PhysRevB.54.11169}.

\bibitem[{\citenamefont{Perdew et~al.}(1996)\citenamefont{Perdew, Burke, and
  Ernzerhof}}]{perdew1996generalized}
\bibinfo{author}{\bibfnamefont{J.~P.} \bibnamefont{Perdew}},
  \bibinfo{author}{\bibfnamefont{K.}~\bibnamefont{Burke}}, \bibnamefont{and}
  \bibinfo{author}{\bibfnamefont{M.}~\bibnamefont{Ernzerhof}},
  \bibinfo{journal}{Physical review letters} \textbf{\bibinfo{volume}{77}},
  \bibinfo{pages}{3865} (\bibinfo{year}{1996}).

\bibitem[{\citenamefont{Kresse and Joubert}(1999)}]{PhysRevB.59.1758}
\bibinfo{author}{\bibfnamefont{G.}~\bibnamefont{Kresse}} \bibnamefont{and}
  \bibinfo{author}{\bibfnamefont{D.}~\bibnamefont{Joubert}},
  \bibinfo{journal}{Phys. Rev. B} \textbf{\bibinfo{volume}{59}},
  \bibinfo{pages}{1758} (\bibinfo{year}{1999}),
  \urlprefix\url{https://link.aps.org/doi/10.1103/PhysRevB.59.1758}.

\bibitem[{\citenamefont{Zhou et~al.}(2016)\citenamefont{Zhou, Zhang, Wang, and
  Zhang}}]{zhou2016phonon}
\bibinfo{author}{\bibfnamefont{H.}~\bibnamefont{Zhou}},
  \bibinfo{author}{\bibfnamefont{G.}~\bibnamefont{Zhang}},
  \bibinfo{author}{\bibfnamefont{J.-S.} \bibnamefont{Wang}}, \bibnamefont{and}
  \bibinfo{author}{\bibfnamefont{Y.-W.} \bibnamefont{Zhang}},
  \bibinfo{journal}{Physical Review E} \textbf{\bibinfo{volume}{94}},
  \bibinfo{pages}{052123} (\bibinfo{year}{2016}).

\bibitem[{\citenamefont{Lee and Hwang}(2017)}]{lee2017molecular}
\bibinfo{author}{\bibfnamefont{Y.}~\bibnamefont{Lee}} \bibnamefont{and}
  \bibinfo{author}{\bibfnamefont{G.~S.} \bibnamefont{Hwang}},
  \bibinfo{journal}{Journal of Physics D: Applied Physics}
  \textbf{\bibinfo{volume}{50}}, \bibinfo{pages}{494001}
  (\bibinfo{year}{2017}).

\bibitem[{\citenamefont{Guo and Huang}(2015)}]{guo2015approaching}
\bibinfo{author}{\bibfnamefont{R.}~\bibnamefont{Guo}} \bibnamefont{and}
  \bibinfo{author}{\bibfnamefont{B.}~\bibnamefont{Huang}},
  \bibinfo{journal}{Scientific reports} \textbf{\bibinfo{volume}{5}},
  \bibinfo{pages}{9579} (\bibinfo{year}{2015}).

\bibitem[{\citenamefont{Grosso}(2003)}]{grosso2003giuseppe}
\bibinfo{author}{\bibfnamefont{G.}~\bibnamefont{Grosso}},
  \bibinfo{journal}{Solid State Physics, 2rd edn.(Academic Press, 2013)}
  (\bibinfo{year}{2003}).

\bibitem[{\citenamefont{Allen et~al.}(1999)\citenamefont{Allen, Feldman,
  Fabian, and Wooten}}]{allen1999diffusons}
\bibinfo{author}{\bibfnamefont{P.~B.} \bibnamefont{Allen}},
  \bibinfo{author}{\bibfnamefont{J.~L.} \bibnamefont{Feldman}},
  \bibinfo{author}{\bibfnamefont{J.}~\bibnamefont{Fabian}}, \bibnamefont{and}
  \bibinfo{author}{\bibfnamefont{F.}~\bibnamefont{Wooten}},
  \bibinfo{journal}{Philosophical Magazine B} \textbf{\bibinfo{volume}{79}},
  \bibinfo{pages}{1715} (\bibinfo{year}{1999}).

\end{thebibliography}

\end{document}